# Illustrating $\Delta E = c^2 \Delta m$

Bernhard Rothenstein, Timisoara, Str.Drubeta 32 Romania
brothenstein@gmail.com

*Abstract. We present a thought experiment that involves a high-sensitivity balance initially in a state of mechanical equilibrium but not in a state of thermal equilibrium. After reaching the state of thermal equilibrium it is no longer in a state of mechanical equilibrium.*

Antippa[1] presents a series of derivations convincing us that, when speaking about a photon, we should characterize it by its energy $E_{ph}$, momentum $p_{ph}$ and mass $m_{ph}$, related by

$$p_{ph} = \frac{E_{ph}}{c} \qquad (1)$$

$$m_{ph} = \frac{E_{ph}}{c^2}. \qquad (2)$$

Einstein's derivation[2] involves a cylindrical box of mass $M$ and a photon of energy $E_{ph}$. Conservation of the centre of mass obliges us to consider that we should associate to the energy $E_{ph}$ a mass $m_{ph}$ related by (2) that is independent of the mass $M$ of the box. Because $m_{ph}$ is associated with the concept of inertia, it is considered that (2) determines the quantitative inertia of energy.

Because an electromagnetic wave propagating in empty space carries photons and energy proportional to the number of photons being carried, we can consider that the energy $E$, the momentum $p$ and the mass $m$ it carries are related by

$$p = \frac{E}{c} \qquad (3)$$

$$m = \frac{E}{c^2} \qquad (4)$$

These equations are confirmed by the fact that the photon heats and exerts pressure on a body on which it is incident. We can consider that (3) and (4) are a direct consequence of classical electrodynamics.[3]

According to Einstein,[4] mass and energy are different manifestations of the same thing, i.e. energy has mass and, conversely, that mass can be thought of as being a form of energy. As a consequence, if a hot body loses an amount of energy **E** in cooling, it also looses an amount of mass $\frac{E}{c^2}$. Conversely, if we heat a body, its mass increases. In order to illustrate these facts, consider a high-sensitivity balance located in a uniform vertical gravitational field. In its left pan it contains a spherical blackbody of mass **$m_i$** heated to a temperature **T** which is higher then the ambient temperature, **$T_0$.** In its right pan we find a body of equal mass **$m_i$**. Under such conditions the balance is in a state of mechanical equilibrium, but not in state of thermal equilibrium. The thought experiment shows us that when the temperature of the blackbody becomes **$T_0$,** the balance reaches a state of thermal equilibrium, but is no longer in the state of mechanical equilibrium. In order to establish a new state of mechanical equilibrium, we should remove a mass $\Delta \mathbf{m}$ from the right pan, in such a way that

$$\Delta \mathbf{E} = \mathbf{c}^2 \Delta \mathbf{m}, \tag{5}$$

where $\Delta \mathbf{E}$ stands for the energy radiated by the blackbody during its cooling. In order to evaluate $\Delta \mathbf{E}$ we should take into account that the cooling takes place with variable mass, i.e.

$$\Delta \mathbf{E} = \mathbf{m}_i \mathbf{C} \mathbf{T} - (\mathbf{m}_i - \Delta \mathbf{m}) \mathbf{C} \mathbf{T}_0. \tag{6}$$

We can consider that our thought experiment is an illustration of an experiment described recently by Rainville et all.[5]

If in the left pan of the balance we would consider a perfectly inelastic collision between two identical lumps of putty having equal and opposite velocities which coalesce into a single lump at rest, then if the balance was in equilibrium before collision, after collision it is not and in order to bring it into a state of equilibrium, we should add to the right pan a mass

$$\Delta \mathbf{m} = \frac{\mathbf{E}_k}{\mathbf{c}^2} \tag{7}$$

where **$E_k$** represents the total kinetic energy converted into heat, heating the compound particle and thus increasing its temperature.[6] The balance is not in a state of thermal equilibrium. After its cooling, we should remove the additional mass in order to establish mechanical equilibrium. The thought experiments presented above do not involve special relativity. They illustrate the important fact that $\Delta \mathbf{E} = \mathbf{c}^2 \Delta \mathbf{m}$ applies to any physical process in which the energy of a body changes with $\Delta \mathbf{E}$ and its mass changes with $\Delta \mathbf{m}$.